\documentclass[prb,twocolumn,showpacs,amsmath,amssymb]{revtex4}
\usepackage{graphicx}       
\usepackage{dcolumn}        
\newcommand{\beq}{\begin{equation}}
\newcommand{\eeq}{\end{equation}}
\newcommand{\beqa}{\begin{eqnarray}}
\newcommand{\eeqa}{\end{eqnarray}}
\newcommand{\kvec}{{\bf k}}
\newcommand{\qvec}{{\bf q}}
\let\Re\relax\DeclareMathOperator{\Re}{Re}
\let\Im\relax\DeclareMathOperator{\Im}{Im}
\newcommand{\abs}[1]{\left|#1\right|}
\begin{document}
\title{Optical conductivity near finite-wavelength quantum criticality}
\author{S. Caprara, M. Grilli, C. Di Castro, and T. Enss} 
\affiliation{SMC-INFM-CNR Rome and Dipartimento di Fisica\\
Universit\`a di Roma ``La Sapienza'', Piazzale Aldo Moro 5, I-00185 Roma, 
Italy}
\begin{abstract}
We study the optical conductivity $\sigma(\Omega)$ of an electron system near 
a quantum critical point with finite-wavelength ordering. $\sigma(\Omega)$ 
vanishes in clean Galilean-invariant systems, unless electrons are 
coupled to dynamical collective modes, which dissipate the current. 
This coupling introduces a non-universal energy scale. Depending on the 
parameters of each specific system, a variety of responses arise near 
criticality: scaling peaks at a temperature- and 
doping-dependent frequency, peaks at a fixed frequency, or no peaks to be 
associated with criticality. Therefore the lack of scaling in the far-infrared
conductivity in cuprates does not necessarily call for new concepts of 
quantum criticality.
\end{abstract}
\date{\today}
\maketitle
Quantum criticality has become a central issue in solid-state physics and 
might account for the non-Fermi-liquid properties of 
cuprates \cite{chubukov,CDG,varma,pomeranchuk}, heavy fermions \cite{coleman} 
and other strongly correlated electron systems. A relevant goal is to 
determine the role of critical fluctuations in various anomalous properties. 
In this context, controversial results on the scaling behavior of
optical conductivity $\sigma(\Omega)$ have been found
\cite{vandermarel,lupi,singley,lucarelli,dumm,bernhard,timusk}.
The issues of scaling violations and the possible need of 
``concepts beyond the standard model of 
quantum criticality'' \cite{vandermarel} have been discussed on the basis both 
of experimental \cite{vandermarel} and theoretical \cite{phillips} arguments. 
In this Rapid Communication we consider the optical conductivity of a system near a quantum 
critical point (QCP). We show that difficulties in reconciling optical 
conductivity experiments with the standard quantum criticality framework does not
necessarily call for new concepts of quantum criticality, but may naturally 
arise from the specific nature of this response function: $\sigma(\Omega)$ 
measures absorption and requires a mechanism for 
current dissipation (impurities, umklapp processes, dynamical phonons all 
play this role in real systems). This mechanism, in turn, involves additional 
energy scales possibly 
leading to violation of scaling in some parameter range. 

Our analysis is carried 
out for a neutral order parameter with critical collective  modes (CM's) 
at finite wave vectors $\qvec_c$. This is not only pertinent to the cuprates,
which are the main object of this work and where a relevant role of spin-ordering 
\cite{chubukov} and/or charge-ordering \cite{CDG} fluctuations has been 
proposed, but may apply to some heavy fermions, dichalcogenides, and other 
systems where spatial (usually spin or charge) order occurs at low 
temperature. The {\it neutral} order parameter fluctuations 
couple in a non-trivial dynamical way with the external electromagnetic field. 

We address two issues: (i) to what extent critical CM's, usually characterized 
by a strong dependence on temperature and other control parameters (like 
pressure or chemical doping $x$), produce specific signatures 
in $\sigma(\Omega)$ and (ii) how dissipation mechanisms introduce additional 
scales, which may mask the quantum critical features of optical spectra 
in some parameter range. 

In the first part of this Rapid Communication we consider the textbook case of 
Galilean-invariant
electrons, where current dissipation coincides with momentum dissipation.
We will later see that the Ward identities enforcing current conservation 
are in fact
more general when implemented for the leading contribution arising from 
scattering due to CM's at finite $\qvec_c$. In this case our results hold
true for electrons in a lattice whenever the umklapp processes 
for critical scattering are inactive.

\begin{figure}
\includegraphics[scale=0.23]{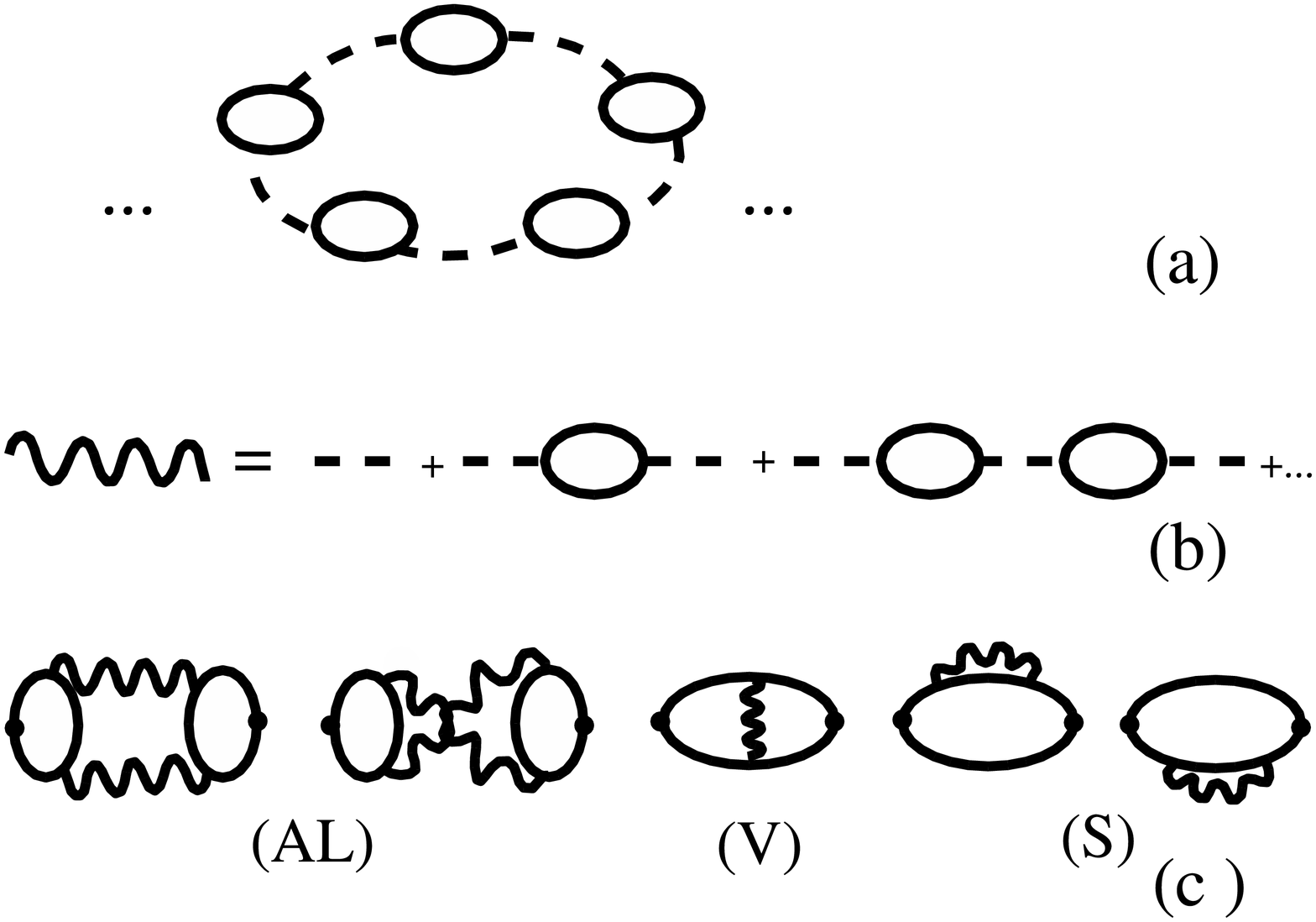}
\vspace{-0.4truecm}
\caption{(a) Typical diagram of our BK generating functional. The solid and 
dashed lines represent the QP propagator and the bare interaction $V(q)$, 
respectively. (b) Bubble resummation to obtain the RPA-dressed CM propagator 
(wavy line). (c) Diagrams for the current-current response, obtained from 
current-vertex insertions (solid dots) in the diagrams of the BK functional. 
 AL, V, and S stand for Aslamazov-Larkin-like, 
vertex, and self-energy diagrams, respectively.}
\label{fig.1}
%
\end{figure}

{\it --- The model ---} We start from a model effective action at
temperature $T$
\begin{eqnarray*}
{\cal S}&=& T\sum_{k,\sigma}G_0^{-1}(k) c^\dagger_{k,\sigma}c_{k,\sigma}+
T\sum_q\chi_0^{-1}(q) \phi_q\phi_{-q} \nonumber \\
&+& T^2 \sum_q\sum_{k,\sigma,\sigma'} 
\gamma_{\sigma \sigma'}c^\dagger_{k+q,\sigma}c_{k,\sigma'}\phi_{-q},\nonumber
\end{eqnarray*}
where $k\equiv(\kvec,\epsilon_\ell)$, $q\equiv(\qvec,\omega_n)$,  
$\epsilon_\ell$ $(\omega_n)$ are fermionic (bosonic) Matsubara frequencies, 
$G_0=( i\epsilon_\ell-\xi_\kvec)^{-1} $ is the bare quasiparticle (QP) 
propagator, $\xi_\kvec$ is the dispersion of the QP's created by the $c$ fields, and 
$\gamma$ is the (theory-dependent) vertex coupling the QP's to collective 
charge or spin excitations, represented by the bosonic fields $\phi$. 
$\chi_0$ is the bare CM propagator, i.e., the 
charge or spin susceptibility in the absence of QP polarization
dressing. At the bare level the 
$\phi$ fields mediate an effective electron-electron interaction 
$V(q)=\gamma^2\chi_0(q)$. If this interaction is purely static, it cannot lead 
to electron momentum dissipation in a Galilean-invariant system
and, as we show below, $\sigma(\Omega)$ 
vanishes at any finite frequency, provided it is properly calculated within a 
conserving perturbative scheme. If instead this bare interaction has its own 
dynamics arising, e.g., from phonons \cite{CDG,eggbox}, electron momentum is 
dissipated and the optical response is finite. For the sake of concreteness we 
consider the charge case, where the $\phi$ fields and the $\gamma$ vertex are 
scalar. To make 
contact with the case of dichalcogenides and cuprates, we consider a model 
in two dimensions. We perform a conserving calculation starting from the 
simplest bubble-type Baym-Kadanoff (BK) functional \cite{baym}, 
which involves only bare QP propagators and the bare interaction $V(q)$  [see 
 Fig.\ \ref{fig.1}(a)]. Once current vertices are inserted, to calculate 
$\sigma(\Omega)$ by means of the Kubo formula, the bubble diagrams can be 
resummed by introducing a random-phase-approximation- (RPA) dressed CM propagator [see Fig.\ 
\ref{fig.1}(b)]. Then, the full set of conserving diagrams reported in Fig.\ 
\ref{fig.1}(c) is obtained. These diagrams give the CM corrections to the bare 
current-current response. In a Galilean-invariant system, $\sigma(\Omega)$ maintains 
a $\delta$-like Drude term, while the critical CM's may 
give absorption at finite  frequencies. 

The RPA effective interaction 
of Fig.\ \ref{fig.1}(b), $\Gamma=(V^{-1}-\Pi)^{-1}$ [here 
$\Pi(q)\equiv -T\sum_k G_0(k+q)G_0(k)$ is the QP
polarization bubble], near the QCP has the generic form
\beq
\Gamma(q)=-{\Pi_\omega^{-1}}\left(m+\varepsilon_\qvec+\abs{\omega_n}+
\omega_n^2/{\overline \Omega}\right)^{-1},\label{dynamprop}
\eeq
for small frequencies and $\qvec\approx\pm \qvec_c$. Here, $\Pi_\omega\equiv
[\Pi(\qvec_c,\omega_n)-\Pi(\qvec_c,0)]/\abs{\omega_n} \vert_{\omega_n=0}$,
and $\varepsilon_\qvec\approx\nu |\qvec \mp \qvec_c|^2$. 
$m$, $\nu$, and $\overline \Omega$ are model-dependent coefficients
which result from the second-order expansion of $V^{-1}-\Pi$ around $\qvec_c$ 
and $\omega_n=0$. Except for the 
$\omega_n^2$ term, $\Gamma(q)$ has the general hydrodynamic form of a propagator for 
diffusive CM's (damped by QP's) near a Gaussian QCP. Here $m$, proportional to the 
square of the inverse correlation length, is the CM mass measuring the distance from 
criticality. If the frequency dependence of $\Gamma$ 
only arises from the QP bubble $\Pi$, i.e., if the bare interaction $V$ is 
purely static, the optical response is zero [the diagrams of Fig.\ 1(c) cancel 
each other], as we show below. On the other hand, if the bare interaction $V$ 
has its own dynamics, a finite $\sigma(\Omega)$ is obtained. 
 We analyze these two situations, and for concreteness we 
 fix the values of the parameters,
adopting as an example the model of Ref.\ \onlinecite{eggbox}.  There, 
a bare interaction 
$V(q)=V_0(\qvec)-\lambda\bar\omega^2(\bar\omega^2+\omega_n^2)^{-1}$ was 
considered, arising from both a static short- and long-range Coulomb 
repulsion $V_0$, and from the coupling to a dispersionless phonon of frequency 
$\bar\omega$.  For moderate  electron-phonon coupling
$\lambda \lesssim \varepsilon_F$, where $\varepsilon_F$ 
is the typical QP Fermi energy  (in 
cuprates, e.g., $\varepsilon_F\approx 0.3$ eV), this interaction can lead to a 
charge-ordering instability, at a wave vector $\qvec_c$.
Within this model we find $m\equiv\Pi_\omega^{-1}\{[\lambda-V_0(\qvec_c)]^{-1}
+\Pi(\qvec_c,0)\}$, and ${\overline\Omega}\equiv\lambda^{-1}\bar\omega^2
[\lambda-V_0(\qvec_c)]^2 \Pi_\omega$.  In Refs.\ 
\onlinecite{CDG} and \onlinecite{eggbox} the instability, signaled by a vanishing $m$, occurred for 
$\lambda\sim V_0\sim\abs{\Pi(\qvec_c,0)}^{-1}\sim \varepsilon_F$.  
Then, we estimate $\nu\sim  \varepsilon_F/k_F^2$ ($k_F$ is the Fermi momentum), 
and ${\overline\Omega}\sim\bar\omega(\bar\omega/ \varepsilon_F)$.

\begin{figure}
\includegraphics[scale=0.28]{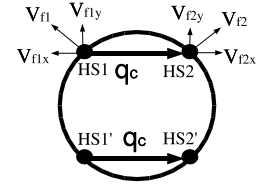}
\vspace{-0.4truecm}
\caption{Sketch of a two-dimensional Fermi surface with four HS's connected by 
a generic incommensurate critical wave vector $\qvec_c=(q_c,0)$. Fermi 
velocities at the HS's and their components are also shown.}
\label{fig.2}
%
\end{figure}

{\it --- Current-current response function ---}
Near criticality, the diagrams of Fig.\ \ref{fig.1}(c), with incoming zero 
momentum and finite frequency $\Omega_l$, are dominated by the poles of the 
CM propagators. Then, we write the vertex--self-energy (VS) and Aslamazov-Larkin (AL) contributions to the 
current-current response function $\chi_{jj}^{\alpha\alpha}$ as 
$T\sum_{\qvec,\omega_n} {\mathcal V}_{VS}^{\alpha\alpha}(\omega_n,\Omega_l) 
\Gamma(\qvec,\omega_n)$ and $\frac{1}{2}T\sum_{\qvec,\omega_n} 
[{\mathcal V}_{AL}^\alpha(\omega_n,\Omega_l)]^2\Gamma(\qvec,\omega_n)
\Gamma(\qvec,\Omega_l+\omega_n)$, where $\alpha=x,y$ and 
we have exploited the relation
$\Gamma(\qvec,\omega_n)=\Gamma(-\qvec,\omega_n)$. The vertices 
${\mathcal V}_{VS}^{\alpha\alpha}$ and ${\mathcal V}_{AL}^\alpha$ come 
from the integration on the QP loops. To perform an analytic calculation, we 
adopt the standard procedure of linearizing the QP dispersion around the 
points of the Fermi surface connected by $\qvec_c$ [hot spots (HS's)]
\cite{chubukov}. Then, we find (see also Ref.\ \onlinecite{TS})
\begin{eqnarray}
{\mathcal V}_{VS}^{\alpha\alpha}(\omega_n,\Omega_l)&=& 
-\frac{e^2}{2} \Pi_\omega(u^\alpha)^2\Omega_l^{-2}\nonumber\\
~&\times&\left(\abs{\Omega_l+\omega_n} -2\abs{\omega_n}+
\abs{\Omega_l-\omega_n}\right), \label{vvs}\\
{\mathcal V}_{AL}^\alpha(\omega_n,\Omega_l)&=&
-i e \Pi_\omega u^\alpha{\Omega_l^{-1}}
(\abs{\Omega_l+\omega_n}-\abs{\omega_n}),\label{val}
\end{eqnarray}
where $e$ is the electron charge, 
$u^\alpha\equiv v_{HS1}^{\alpha}-v_{HS2}^{\alpha}$, and $v_{HS}^\alpha$ is the 
$\alpha$ component of the Fermi velocity at the HS's (see Fig.\ \ref{fig.2}). 
The vertices vanish identically in the direction perpendicular to $\qvec_c$, 
i.e., for $\alpha=y$, since in Fig.\ \ref{fig.2} we took $\qvec_c$ along the 
$x$ axis and $v_{HS1}^y=v_{HS2}^y$, i.e., $u^y\equiv 0$. Thus 
$\chi_{jj}^{yy}(\Omega_l)\equiv 0$ for $\Omega_l\neq 0$, regardless of the 
retarded or static character of the bare interaction $V$.

To calculate $\chi_{jj}^{xx}$, we exploit the identity
\begin{equation}
\Gamma(\qvec,\omega_n)\Gamma(\qvec,\Omega_l+\omega_n)=
\frac{\Pi_\omega^{-1}
[\Gamma(\qvec,\Omega_l+\omega_n)-\Gamma(\qvec,\omega_n)]}
{\abs{\Omega_l+\omega_n}-\abs{\omega_n}
+\frac{\Omega_l}{\overline\Omega}(\Omega_l+2\omega_n)},\label{wi}
\end{equation}
which is immediately derived from Eq.\ (\ref{dynamprop}), and allows us to
write the AL contribution to $\chi_{jj}^{xx}$ as
\[
-T\sum_{\qvec,\omega_n} 
\frac{\Pi_\omega^{-1}[{\mathcal V}_{AL}^x(\omega_n,\Omega_l)]^2}
{\abs{\Omega_l+\omega_n}-\abs{\omega_n}
+\frac{\Omega_l}{\overline\Omega}(\Omega_l+2\omega_n)}\Gamma(\qvec,\omega_n),
\]
where a single effective interaction appears, at the expense of a more 
complicated prefactor.

It is instructive to consider first the case 
${\overline\Omega}\to\infty$, (i.e., $\bar\omega\to\infty$). In this case the 
QP's are only coupled among themselves and with a static phonon field, and the 
frequency dependence in Eq.\ (\ref{dynamprop}) only comes from the QP bubble. 
As a consequence, $\sigma(\Omega)$ must vanish for any finite frequency, while 
only the $\delta$-like Drude term remains. For ${\overline\Omega}\to\infty$
and by means of Eq. (\ref{val}), Eq.\ (\ref{wi}) can be
cast in the form of a Ward identity $eu^\alpha
[\Gamma(\qvec,\Omega_l+\omega_n)-\Gamma(\qvec,\omega_n)]
=i\Omega_l{\mathcal V}_{AL}^\alpha(\omega_n,\Omega_l)
\Gamma(\qvec,\Omega_l+\omega_n)\Gamma(\qvec,\omega_n)$, which plays the role 
of the ordinary Ward identity exploited, e.g., in Ref. \onlinecite{ambegaokar}, to 
prove the cancellation of the paraconductivity, in the case when the electron 
momentum commutes with the Hamiltonian. In our case, by means of Eqs. 
(\ref{vvs}) and (\ref{val}) we can eliminate the VS vertex in favor of the AL 
vertex, ${\mathcal V}_{VS}^{xx}(\omega_n,\Omega_l)=
-\frac{ie}{2}u^x\Omega_l^{-1}
[{\mathcal V}_{AL}^x(\omega_n,\Omega_l)+
{\mathcal V}_{AL}^x(-\omega_n,\Omega_l)]$.
Then, $\chi_{jj}^{xx}=\frac{ie}{2}u^x\Omega_l^{-1}
T\sum_{\omega_n,\qvec}[{\mathcal V}_{AL}^x(\omega_n,\Omega_l)-
{\mathcal V}_{AL}^x(-\omega_n,\Omega_l)]\Gamma(\qvec,\omega_n)$
vanishes identically, since the quantity between square brackets is an odd 
function of $\omega_n$.

{\it --- Momentum dissipation: finite ${\overline \Omega}$ ---}
To obtain an optical response it is therefore necessary to introduce a 
dissipation mechanism. In this Rapid Communication we implement this dissipation via the 
$\omega_n^2$ term in Eq.\ (\ref{dynamprop}), which in the example
we are considering arises from the phonon 
dynamics. In this case, after analytic continuation of the external Matsubara 
frequency, we find
\beqa
\chi_{jj}^{xx}(\Omega)&=&
\frac{A}{(i\Omega-{\overline\Omega})\Omega} 
P\int_{-\infty}^{+\infty} \frac{dz}{i\pi}
\ln\left[\frac{z^2-\overline\Omega(\Lambda-iz)}
{z^2-\overline\Omega(m-iz)}\right]\nonumber \\
&\times& \left[ \frac{\overline\Omega\Omega+i(2z+\Omega)(z-\Omega)}
{{\overline \Omega}-i(2z+\Omega)}\mathrm{coth}
\left(\frac{z}{2T}\right)\right.\nonumber \\
&-& \left. \frac{i(2z-\Omega)(z-\Omega)}
{{\overline \Omega}-i(2z-\Omega)} 
\mathrm{coth}\left(\frac{z-\Omega}{2T}\right)\right],\label{finalchi}
\eeqa
where $A=e^2 (v_{HS}^x)^2/\nu d$ is the dimensional prefactor, $d$ is the 
interlayer distance which translates the two-dimensional response into the 
in-plane response of a layered system, and $\Lambda$ is an ultraviolet cutoff. 
Extracting the conductance quantum  $e^2/h$ and the factor $1/d$, we are left 
with a dimensional factor which we estimate as $h(v_{HS}^x)^2/\nu \sim 
\varepsilon_F/h$.

One can check that $\Im\chi_{jj}^{xx}(\Omega)$ linearly vanishes with 
$\Omega$, giving a finite $\sigma(\Omega)=\Im\chi_{jj}^{xx}(\Omega)/\Omega$, 
for $\Omega\to 0$. However, in our calculation without disorder, a 
$\delta$-like Drude term $[D_0-\pi \Re\chi_{jj}^{xx}(0)]\delta(\Omega)$ is 
still present, where $D_0$ is the QP Drude weight in the absence of the CM 
contribution. Kramers-Kronig relations connecting $\Re\chi_{jj}^{xx}$ and 
$\Im\chi_{jj}^{xx}$ guarantee spectral weight conservation within our 
conserving approach: the finite-frequency weight associated with 
$\sigma(\Omega)$ is exactly subtracted from $D_0$ by 
$\pi \Re\chi_{jj}^{xx}(0)$, i.e., 
$\int d\Omega~ \left\{ \left[ D_0-\pi \Re\chi_{jj}^{xx}(0)\right]\delta(\Omega)
+\sigma(\Omega)\right\}=D_0$.

In Eq.\ (\ref{finalchi}), the ${\overline\Omega}$ dependence is crucial and 
introduces a non-critical energy scale in the absorption, which may strongly 
alter the dependence of the optical spectra on the CM mass. The scale 
${\overline \Omega}$ (which we estimated for the cuprates as a rather low 
energy scale $\sim\bar\omega^2/ \varepsilon_F$) 
determines, at low $T$, the frequency below 
which the response tends to be vanishingly small, as it is for 
${\overline \Omega}=\infty$. On the other hand, for 
$\Omega\gtrsim{\overline \Omega}$ a finite absorption is found. We discuss a 
system in the quantum critical regime where the mass 
$m(T,x_{QCP})\equiv m(T)=\alpha T$.\cite{andergassen}

\begin{figure}
\includegraphics[scale=0.6]{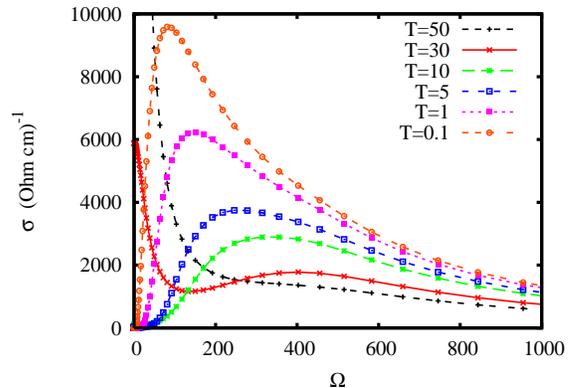}
\vspace{-0.4truecm}
\caption{(Color online) Optical conductivity for a system of QP's coupled to 
CM's in the quantum-critical regime $(x=x_{QCP})$, with $m=30T$ and 
${\overline \Omega}=30$ cm$^{-1}$. All frequencies and temperatures are in 
cm$^{-1}$. The interlayer distance was taken as $d=10^{-9}$ m.}
\label{fig.3}
%
\end{figure}

\begin{figure}
\includegraphics[scale=0.6]{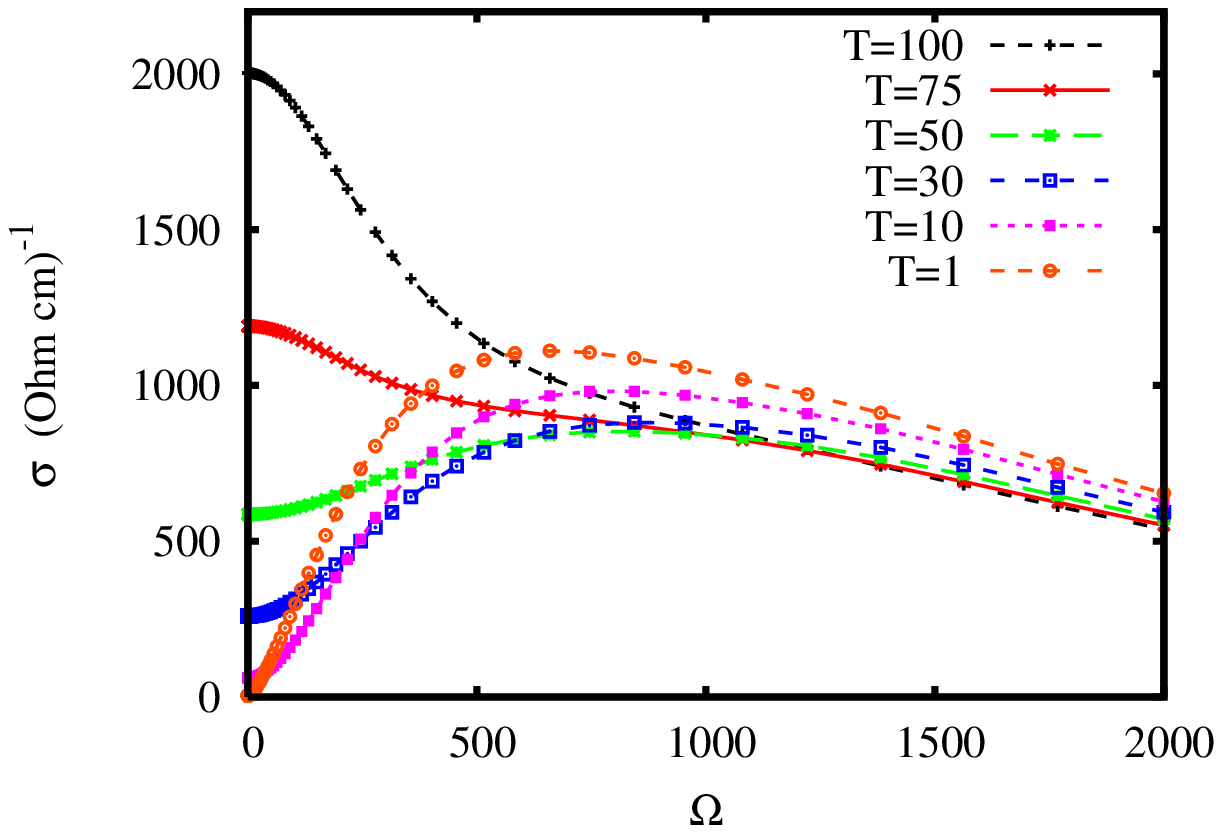}
\vspace{-0.4truecm}
\caption{(Color online) Same as Fig.\ \ref{fig.3}, but with $m=3T$ and  
${\overline \Omega}=330$ cm$^{-1}$.}
\label{fig.4}
%
\end{figure}

We first consider the case $\alpha=30$ and ${\overline \Omega}=30$ cm$^{-1}$, 
illustrated in Fig.\ \ref{fig.3}. As long as $m(T)\gtrsim{\overline\Omega}$, 
the peak position is ruled by the critical energy scale $m$ and the strong $T$ 
dependence of the CM mass is visible as  peaks in $\sigma(\Omega)$ at
temperature-dependent frequencies: 
the peaks soften and sharpen upon decreasing $T$. Thermal effects tend to fill 
the spectra below the peaks, which are clearly visible only if the CM mass, 
i.e., the peak frequency, is substantially larger than $T$. We find indeed 
absorption peaks at frequencies $\Omega\sim m(T)$ 
only if the non-universal coefficient $\alpha\gtrsim 10$, as in Fig.\ 
\ref{fig.3}. On the other hand, in Fig.\ \ref{fig.4} we take 
$m(T)=3T \lesssim{\overline\Omega}\approx 330$ cm$^{-1}$ and the peak position 
is fixed at $\overline \Omega$.\cite{footnote} In this case, thermal effects 
show up for $T\gtrsim 0.2{\overline \Omega}$ with absorption at low frequency. 
Then, the peaks are gradually embedded in this thermally-generated absorption. 
This shows that the absence of well-separated absorption peaks in the far 
infrared optical spectra does not necessarily mean that critical CM's 
contributing to $\sigma(\Omega)$ are absent. 

{\it --- Discussion ---} So far we considered  clean Galilean-invariant 
systems. The presence of a lattice introduces two effects:
the current vertex is no longer proportional to the electron momentum and
umklapp scattering processes may occur. At leading order in the critical 
scattering, the first effect is immaterial because the current vertex is evaluated
at the HS's and the velocity only appears in multiplicative prefactors.
Then, for ${\overline \Omega} \to \infty$ the
Ward identities implementing the cancellation remain valid, while Eq.
(\ref{finalchi}) holds for finite ${\overline \Omega}$.
The second effect may instead be significant. However,
our results for the leading critical contributions remain valid 
in those systems where $\qvec_c$ 
cannot connect Fermi surfaces in different Brillouin zones. 
Our analysis also provides reliable results for dirty systems 
in the frequency range above the 
typical impurity scattering rate $1/\tau$. In assessing the results discussed 
above, we strongly relied on a conserving scheme to calculate 
$\sigma(\Omega)$. Although the relevance of vertex corrections was previously 
proposed \cite{KKU}, in this Rapid Communication we provide quantitative evidence that
a conserving scheme captures the strong cancellations
occurring in systems with current conservation and it keeps partial cancellations
also when the current conservation law is not obeyed.
Therefore, non-conserving approaches in models with strong momentum dependence
\cite{casek} should be handled with care.\cite{altri} 

Our main result is that (the leading critical contribution to)
 optical absorption at finite frequency only appears, 
in a clean Galilean-invariant system or in systems without
critical umklapp scattering, when electrons are coupled to 
other dynamical degrees of freedom, which mediate a retarded interaction 
allowing for electron-current dissipation. This introduces a non-universal 
energy scale $\overline\Omega$. If the CM mass $m$, controlling the distance 
from criticality, is larger than ${\overline\Omega}$, the CM peak position is 
ruled by $m$ and follows its $x$ and $T$ dependence. Otherwise, the peak is 
pinned at a frequency $\overline\Omega$. This saturation, which occurs even 
when $m$ vanishes with $T$ $(x=x_{QCP})$, is an intrinsic feature irrespective of 
other mechanisms (e.g., pinning), which would keep $m$ finite. 
Therefore, it is crucial to recognize that the saturation of the peak in far 
infrared spectra cannot generically be used as evidence against criticality.
Thermal filling reduces the visibility of the CM peak to the regimes where 
the peak frequency is substantially larger than $T$. Whenever the above 
conditions are not fulfilled, the CM's provide broad absorption.
In real systems, such absorption would interplay with 
the Drude peak and could be confused with an additional impurity broadening. 

Experiments in cuprates provide a wealth of different behaviors, which may 
find counterparts in the various regimes discussed above. In 
${\rm Bi_2Sr_2CuO_6}$ (Ref.\ \onlinecite{lupi}) and ${\rm Nd_{2-x}Ce_xCuO_{4-y}}$  
(Ref.\ \onlinecite{singley}) peaks at temperature-dependent
frequencies, clearly separated from the Drude peak, are 
observed, with approximate scaling behavior. Within our theory, this implies 
$m>{\overline \Omega}$. Peaks at temperature-independent frequencies
(but still distinct from the 
Drude peak) are found in ${\rm La_2Sr_xCuO_4}$ (Ref.\ \onlinecite{lucarelli}), 
${\rm La_{1.6-x}Nd_{0.4}Sr_xCuO_4}$ (Ref.\ \onlinecite{dumm}), and some 
${\rm YBa_2Cu_3O_{7-y}}$ (Ref.\ \onlinecite{bernhard}) samples. This behavior might be due
to a non-vanishing mass or to a saturation at ${\overline \Omega}$. No peaks 
at all are observed in ${\rm Bi_2Sr_2Cu_2O_8}$ (Ref.\ \onlinecite{vandermarel}) and 
${\rm YBa_2Cu_3O_{7-y}}$ (Ref.\ \onlinecite{timusk}). Of course, an explicit 
quantitative analysis of the optical conductivity at low frequencies
requires a proper treatment of disorder and umklapp processes, which is presently in 
progress. Nevertheless our theory accounts for the variety of experimental
behaviors and may reconcile standard quantum criticality with the lack
of universal scaling of the conductivity.

We acknowledge interesting discussions with C. Castellani and J. Lorenzana
and financial support from the MIUR-PRIN 2005, prot.\ 2005022492
and from the Alexander von Humboldt foundation.


\begin{thebibliography}{99}
\bibitem{footnote} This saturation always occurs in the quantum critical 
regime. However, in Fig.\ \ref{fig.3}, owing to the low value of 
$\overline \Omega$, the saturation regime is only found at very low $T$.
\end{thebibliography}
\end{document}